\pdfoutput=1
\documentclass{article}

\usepackage[preprint]{neurips_2023}
\usepackage[utf8]{inputenc} 
\usepackage[T1]{fontenc}    
\usepackage{hyperref}       
\usepackage{url}            
\usepackage{booktabs}       
\usepackage{amsfonts}       
\usepackage{nicefrac}       
\usepackage{microtype}      
\usepackage{xcolor}         

\usepackage{microtype}
\usepackage{graphicx}
\usepackage{subfigure}
\usepackage{booktabs} 

\usepackage{hyperref}
\usepackage{url}
\usepackage{graphicx}
\usepackage{makecell}
\usepackage{multirow}
\usepackage{enumitem}
\usepackage{bbm}
\usepackage{color,xcolor}
\usepackage{booktabs}
\usepackage{lineno}
\usepackage{array}
\usepackage{ulem}
\usepackage{makecell}
\usepackage{multicol}
\usepackage{amsmath,amsthm,amsfonts,amssymb,bm}
\usepackage{stfloats}
\usepackage[font=small,labelfont=bf]{caption}
\usepackage{tabulary}
\usepackage{algorithm}
\usepackage{algorithmicx}
\usepackage{algpseudocode}
\usepackage{bbding}
\usepackage{natbib}
\newlist{compactenum}{enumerate}{4}
\setlist[compactenum,1]{nolistsep}
\newcolumntype{C}[1]{>{\centering\arraybackslash}p{#1}}
\newcolumntype{L}[1]{>{\arraybackslash}p{#1}}

\newcommand{\sigmoid}{\sigma}
\newcommand\NoThen{\renewcommand\algorithmicthen{}}

\newcommand\NoDo{\renewcommand\algorithmicdo{}}

\algnewcommand{\algorithmicendif}{}
\algnewcommand{\algorithmicforend}{}

\algtext*{EndIf}
\algtext*{EndFor}

\author{Zhangyang Gao$^{\dagger}$, Cheng Tan$^{\dagger}$, Stan Z. Li$^{*}$\\
AI Lab, Research Center for Industries of the Future, Westlake University \\
\texttt{\{gaozhangyang, tancheng,Stan.ZQ.Li\}@westlake.edu.cn}\\
\thanks{$^{\dagger}$Equal Contribution, $^{*}$Corresponding Author.}
}

\makeatletter
\def\thanks#1{\protected@xdef\@thanks{\@thanks
        \protect\footnotetext{#1}}}
\makeatother

\usepackage[utf8]{inputenc} 
\usepackage[T1]{fontenc}    
\usepackage{hyperref}       
\usepackage{url}            
\usepackage{booktabs}       
\usepackage{amsfonts}       
\usepackage{nicefrac}       
\usepackage{microtype}      
\usepackage{xcolor}         
\usepackage{wrapfig}

\title{Knowledge-Design: Pushing the Limit of Protein Design via Knowledge Refinement}

\begin{document}
\maketitle

\vspace{-7mm}
\begin{abstract}
  \vspace{-2mm}
  Recent studies have shown competitive performance in protein design that aims to find the amino acid sequence folding into the desired structure. However, most of them disregard the importance of predictive confidence, fail to cover the vast protein space, and do not incorporate common protein knowledge. After witnessing the great success of pretrained models on diverse protein-related tasks and the fact that recovery is highly correlated with confidence, we wonder whether this knowledge can push the limits of protein design further. As a solution, we propose a knowledge-aware module that refines low-quality residues and introduce a memory-retrieval mechanism to save more than 50\% of the training time. We extensively evaluate our proposed method on the CATH, TS50, and TS500 datasets. The experimental results show that Knowledge-Design outperforms the previous PiFold by approximately 9\% on the CATH dataset. Specifically, Knowledge-Design is the first method that achieves 60+\% recovery on CATH, TS50 and TS500 benchmarks. We also provide additional analysis to demonstrate the effectiveness of our proposed method. The code will be publicly available.
\end{abstract}

\vspace{-6mm}
\section{Introduction}
\vspace{-3mm}
Protein sequences, which are linear chains of amino acids, play a crucial role in determining the structure and function of cells and organisms. In recent years, there has been significant interest in designing protein sequences that can fold into desired structures \citep{pabo1983molecular}. Deep learning models \citep{li2014direct, wu2021protein, pearce2021deep, ovchinnikov2021structure, ding2022protein, gao2020deep, gao2022alphadesign, dauparas2022robust, ingraham2019generative, jing2020learning, tan2022generative, hsu2022learning, o2018spin2, wang2018computational, qi2020densecpd, strokach2020fast, chen2019improve, zhang2020prodconn, huang2017densely, anand2022protein, strokach2022deep, li20143d, greener2018design, karimi2020novo,anishchenko2021novo, cao2021fold2seq, liu2022rotamer, mcpartlon2022deep, huang2022accurate, dumortier2022petribert, li2022neural, maguire2021xenet, li2022terminator} have made significant progress in this area. However, many of these methods either ignore the importance of predictive confidence, fail to cover the vast protein space, or lack consideration of common protein knowledge. We argue that the absence of common protein knowledge limits the generalizability of protein design models, and that predictive confidence can help to identify low-quality residues. Therefore, we propose a confidence-aware module that refines low-quality residues using structural and sequential embeddings extracted from pretrained models, thereby generating more rational protein sequences.

\begin{minipage}{\linewidth}
    \begin{minipage}{0.65\linewidth}
        Previous protein design methods have not fully utilized the predictive confidence, i.e., the maximum probability of the residue. Using PiFold \citep{gao2023pifold} as our baseline, we observed significant differences in confidence distributions between positive and negative residues, as shown in Fig. \ref{fig:pifold_distribution}. This finding inspired us to propose a confidence-aware module that automatically identifies low-quality residues and iteratively refines them to reduce prediction errors. However, we encountered a challenge: the recovery rate of PiFold plateaued at around 52\% regardless of how many refine layers (PiGNNs) were added. We hypothesized that this was due to the model being trapped in a local optimum based on the current training set. Escaping the local minimum would require additional inductive bias from other teacher models.
    \end{minipage}
    \hspace{0.00\linewidth}
    \begin{minipage}{0.35\linewidth}
        \begin{figure}[H]
            \centering
            \includegraphics[width=2in]{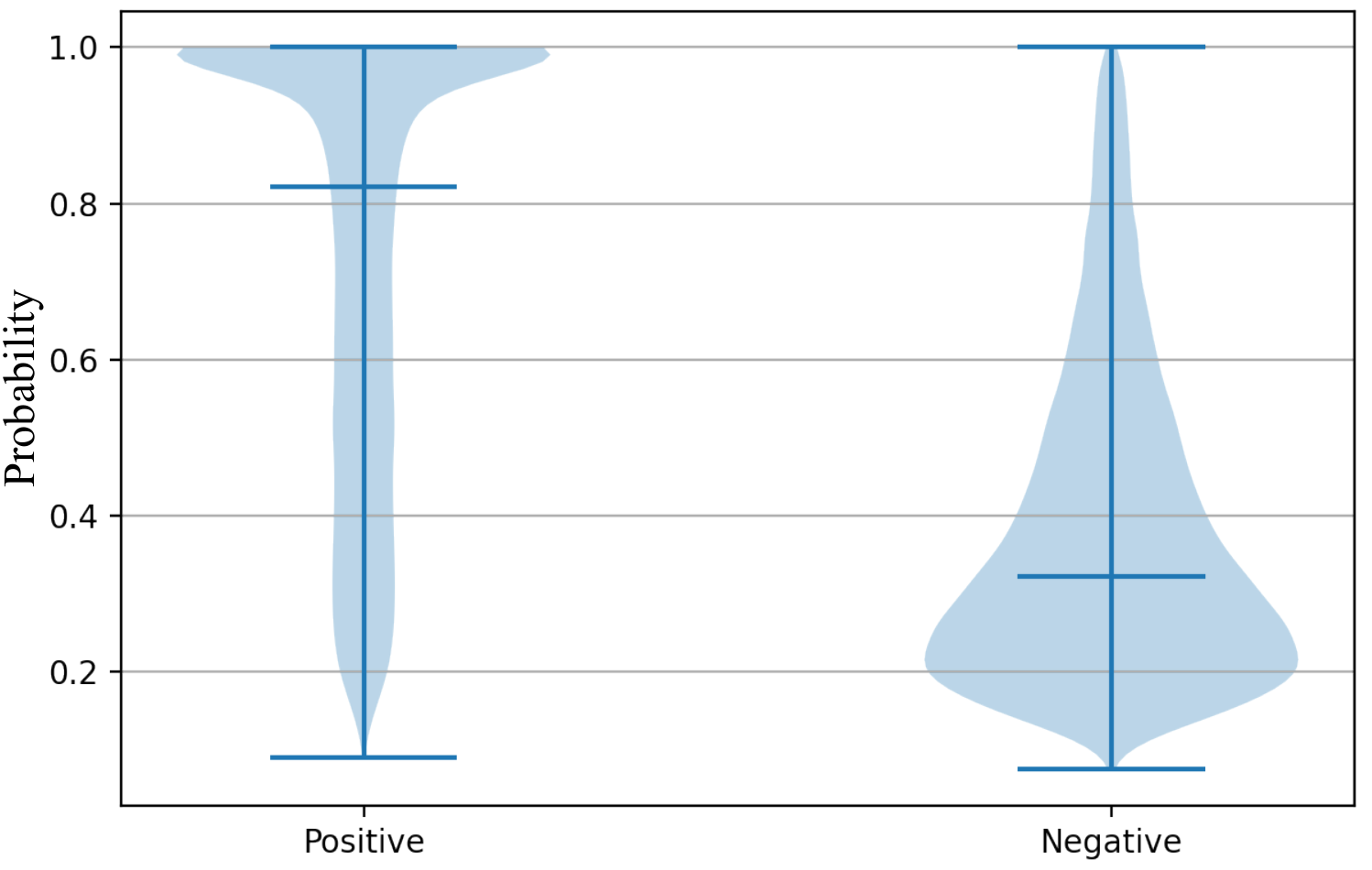}
            \captionsetup{font=small}
            \caption{Confidence of positive and negative residues designed by PiFold\citep{gao2023pifold}. Positive residues are identical to native residues and vice versa.}
            \label{fig:pifold_distribution}
            \vspace{-3mm}
         \end{figure}
    \end{minipage}
\end{minipage}

To escape the local minimum and improve the performance of our protein design model, we suggest leveraging pretrained teacher models. These models have made significant progress on a variety of downstream tasks \citep{zhang2022protein,meier2021language, zhang2022ontoprotein,chen2023data} by learning common knowledge across a vast protein space. The structural knowledge \citep{zhang2022protein, hsu2022learning} can help to learn expressive protein features, while the sequential knowledge \citep{meier2021language} can aid in designing rational proteins. In this study, we investigate three pretrained models, i.e., ESM \citep{meier2021language,lin2022language}, ESM-IF \citep{hsu2022learning}, and GearNet \citep{zhang2022protein}, to extract structural and sequential embeddings as prior knowledge that can enhance our refinement module. As a structure-in and sequence-out task, the structural-based protein design can benefit from the multimodal knowledge and automatically revise residues that violate common sense.

To boost protein design, we propose a confidence-aware refining model that leverages multimodal knowledge. However, we face several challenges: (1) how to adaptively fuse multimodal pretrained knowledge based on the predictive confidence, (2) how to develop more effective refining technologies, and (3) how to efficiently tune the model with large-scale pretrained parameters. Firstly, we propose a multimodal fusion module that combines the knowledge of structure pretraining, sequence pretraining, and history predictions. The predictive confidence is used to control the combination through gated attention, enabling the model to adaptively fuse multimodal knowledge. Secondly, we suggest using virtual MSA and recycling technologies to improve the recovery. Thirdly, we introduce a memory-retrieval mechanism that caches the intermediate results of modules. This mechanism enables the model to retrieve historical embeddings without performing a forward pass, resulting in more than 50\% training time savings.

We call our method Knowledge-Design, a refining methods that considers multimodal knowledge as well as the predictive confidence. We evaluate our method on three benchmark datasets: CATH, TS50, and TS500, and observe significant improvements across all settings. For example, Knowledge-Design is the first method to achieve 60+\% recovery on all three datasets. On the CATH dataset, we observe 9.11\% improvement compared to the previous PiFold method. We also conduct extensive ablation studies to demonstrate how the knowledge-refinement module works and how the memory-retrieval mechanism saves training time. Additionally, we provide further analysis to demonstrate the superiority of our proposed method. Overall, our results demonstrate the effectiveness of our approach in improving protein design performance.

\section{Related work}
Recently, AI algorithms have evolved rapidly in many fields \citep{gao2022simvp, cao2022survey, tan2022simvp, li2022openmixup, he2020momentum, stark2022equibind, gao2023prefixmol}, where the protein folding problem \citep{jumper2021highly,wu2022high,lin2022language,mirdita2022colabfold,wang2022helixfold,li2022uni, gao2023diffsds} that has troubled humans for decades has been nearly solved. Its inverse problem- structure-based protein design - is receiving increasing attention.

\vspace{-2mm}
\paragraph{Problem definition} The structure-based protein design aims to find the amino acids sequence $\mathcal{S} = \{s_i: 1 \leq i \leq n\}$ folding into the desired structure $\mathcal{X}=\{\boldsymbol{x}_{i} \in \mathbb{R}^3: 1 \leq i \leq n \}$, where $n$ is the number of residues and the natural proteins are composed by 20 types of amino acids, i.e., $1\leq s_i \leq 20$ and $s_i \in \mathbb{N}^+$. Formally, that is to learn a function $\mathcal{F}_{\theta}$:
\begin{align}
    \label{eq:protein_dedign}
    \mathcal{F}_{\theta}: \mathcal{X} \mapsto \mathcal{S}.
\end{align}
Because homologous proteins always share similar structures \citep{pearson2005limits}, the problem itself is underdetermined, i.e., the valid amino acid sequence may not be unique \citep{gao2020deep}.

\vspace{-2mm}
\paragraph{MLP-based models} 
MLP is used to predict the probability of 20 amino acids for each residue, and various methods are mainly difficult in feature construction. These methods are commonly evaluated on the TS50, which contains 50 native structures. For example, SPIN \citep{li2014direct} achieves 30\% recovery on TS50 by using torsion angles ($\phi$ and $\psi$), sequence profiles, and energy profiles. Through adding backbone angles ($\theta$ and $\tau$), local contact number, and neighborhood distance, SPIN2 \citep{o2018spin2} improves the recovery to 34\%. Wang's model \citep{wang2018computational} suggests using backbone dihedrals ($\phi$, $\psi$ and $\omega$), the solvent accessible surface area of backbone atoms ($C_{\alpha}, N, C,$ and $O$), secondary structure types (helix, sheet, loop), $C_{\alpha}-C_{\alpha}$ distance and unit direction vectors of $C_{\alpha}-C_{\alpha}$, $C_{\alpha}-N$ and $C_{\alpha}-C$ and achieves 33\% recovery. The MLP method enjoys a high inference speed, but suffers from a low recovery rate because the structural information is not sufficiently considered.

\vspace{-2mm}
\paragraph{CNN-based models} These methods use 2D CNN or 3d CNN to extract protein features \citep{torng20173d, NIPS2017_1113d7a7, NEURIPS2018_488e4104, zhang2020prodconn, qi2020densecpd, chen2019improve} and are commonly evaluated on the TS50 and TS500. SPROF \citep{chen2019improve} adopts 2D CNN to learn residue representations from the distance matrix and achieves a 40.25\% recovery on TS500. 3D CNN-based methods, such as ProDCoNN \citep{zhang2020prodconn} and  DenseCPD \citep{qi2020densecpd}, extract residue features from the atom distribution in a three-dimensional grid box. For each residue, after being translated and rotated to a standard position, the atomic distribution is fed to the model to learn translation- and rotation-invariant features. ProDCoNN \citep{zhang2020prodconn} designs a nine-layer 3D CNN with multi-scale convolution kernels and achieves 42.2\% recovery on TS500. DenseCPD \citep{qi2020densecpd} uses the DensetNet architecture \citep{huang2017densely} to boost the recovery to 55.53\% on TS500. Recent works \citep{anand2022protein} have also explored the potential of deep models to generalize to \textit{de novo} proteins. Despite the improved recovery achieved by the 3D CNN models, their inference is slow, probably because they require separate preprocessing and prediction for each residue.

\vspace{-2mm}
\paragraph{Graph-based models} These methods use $k$-NN graph to represent the 3D structure and employ graph neural networks \citep{defferrard2016convolutional,kipf2016semi,velivckovic2017graph,zhou2020graph,zhang2020deep, gao2022cosp, tan2022target, gao2022semiretro} to extract residue features while considering structural constraints. The protein graph encodes residue information and pairwise interactions as the node and edge features, respectively. GraphTrans \citep{ingraham2019generative} uses the graph attention encoder and autoregressive decoder for protein design. GVP \citep{jing2020learning} proposes geometric vector perceptrons to learn from both scalar and vector features. GCA \citep{tan2022generative} introduces global graph attention for learning contextual features. In addition, ProteinSolver \citep{strokach2020fast} is developed for scenarios where partial sequences are known while does not report results on standard benchmarks. Recently, AlphaDesign \citep{gao2022alphadesign}, ProteinMPNN \citep{dauparas2022robust} and Inverse Folding \citep{hsu2022learning} achieve dramatic improvements. Compared to CNN methods, graph models do not require rotating each residue separately as in CNN, thus improving the training efficiency. Compared to MLP methods, the well-exploited structural information helps GNN obtain higher recovery.

\section{Method}
\subsection{Overall Framework}
\begin{figure}[h]
    \centering
    \includegraphics[width=5.5in]{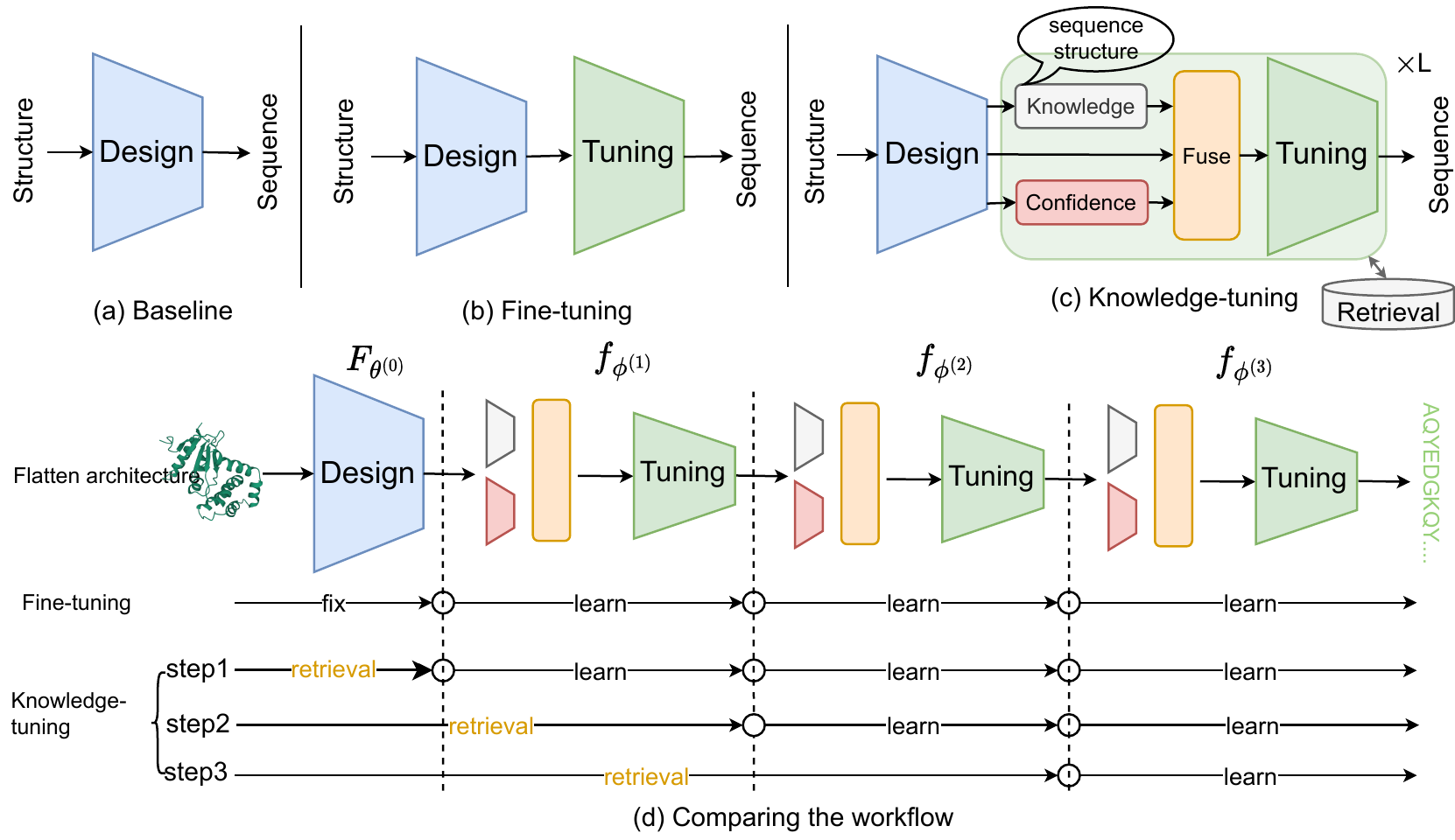}
    \caption{Comparison of various models. (a) The baseline trains from scratch without using pretrained knowledge. (b) The fine-tuning model refines the output of the baseline. (c) The knowledge-tuning model fuse multimodal pretained knowledge and the confidence to enhance the refinement module. 
    (d) The proposed Knowledge-Design model introduces a memory bank to speed up the training process by skipping the forward pass of well-tuned modules.}
    \label{fig:framework}
    \vspace{-3mm}
 \end{figure}

The framework of our Knowledge-Design model is illustrated in Figure \ref{fig:framework}. It comprises an initial design model, denoted by $F_{\theta^{(0)}}$, and $L$ confidence-aware knowledge-tuning modules, denoted by $f_{\phi^{(1)}}, f_{\phi^{(2)}}, \cdots, f_{\phi^{(L)}}$, where $\theta^{(0)}$ and $\phi^{(i)}$ are learnable parameters. To simplify the notation, we write $F_{\theta^{(k)}} = f_{\phi^{(k)}} \cdots \circ f_{\phi^{(1)}} \circ F_{\theta^{(0)}}$, where $\theta^{(k)} = \phi^{(k)} || \cdots || \phi^{(2)} || \phi^{(1)} || \theta^{(0)}$, and $||$ denotes concatenation operation.

For the $l$-th knowledge-tuning module, we denote the protein structure as $\boldsymbol{x} \in \mathbb{R}^{n, 3}$, the residue embedding as $\boldsymbol{h}^{(l)} \in \mathbb{R}^{n, d}$, and the predicted probabilities as $\boldsymbol{p}^{(l)} \in \mathbb{R}^{n, 21}$. Formally, we have $\boldsymbol{h}^{(l)} = f_{\phi^{(l)}} \cdots \circ f_{\phi^{(1)}} \circ F_{\theta^{(0)}}(\boldsymbol{x})$ and $\boldsymbol{p}^{(l)} = \texttt{Predict}^{(l)}(\boldsymbol{h}^{(l)})$. Here, $\circ$ denotes the operation of compositing functions, $n$ is the number of residues, $d$ is the embedding size, $21$ is the number of amino acids plus a special token of [mask], and $\texttt{Predict}^{(l)}()$ is a linear layer equipped with a softmax activation, i.e., $\text{Softmax} \circ \text{Linear}()$. The overall objective of our Knowledge-Design model is to minimize the loss function $\mathcal{L}$ with respect to the learnable parameters $\theta^{(0)}, \phi^{(1)}, \cdots, \phi^{(L)}$:
\begin{align}
\min_{\theta^{(0)}, \phi^{(1)}, \cdots, \phi^{(L)}} \mathcal{L}( f_{\phi^{(L)}} \cdots \circ f_{\phi^{(1)}} \circ F_{\theta^{(0)}}(\boldsymbol{x}) , \boldsymbol{s})
\end{align}
Here, $\boldsymbol{s}$ is the reference sequence, and $\boldsymbol{x}$ is the protein structure.

Under the assumptions that if $\mathcal{L}(F_{\theta^{(k)}}(\boldsymbol{x}), \boldsymbol{s})<\mathcal{L}(F_{\theta^{(k)'}}, \boldsymbol{s})$, then $\mathcal{L}(f_{\phi^{(k+1)}}(F_{\theta^{(k)}}(\boldsymbol{x})), \boldsymbol{s})<\mathcal{L}(f_{\phi^{(k+1)}}(F_{\theta^{(k)'}}(\boldsymbol{x})), \boldsymbol{s})$, indicating that a better initial design model leads to better final results when using the same fine-tuning module, we simplify the objective as:
\begin{align}
    \min_{\phi^{(k)}} \mathcal{L}(f_{\phi^{(k)}}(\boldsymbol{h}^{(k-1)}), \boldsymbol{s}), s.t., \theta^{(k-1)} = \min_{\theta^{(k-1)}} \mathcal{L}(F_{\theta^{(k-1)}}(\boldsymbol{x}), \boldsymbol{s})
    \label{eq:objective_simplified}
\end{align}
Note that $\boldsymbol{h}^{(k-1)} = F_{\theta^{(k-1)}}(\boldsymbol{x}),  k \in \{L, L-1, \cdots, 1\}$. Eq.(\ref{eq:objective_simplified}) suggests the problem could be solved by optimizing the fine-tuning modules $f_{\phi^{(1)}}, f_{\phi^{(2)}}, \cdots, f_{\phi^{(L)}}$ sequentially. Therefore, the parameters $\theta^{(k-1)}$ are frozen when optimizing $\phi^{(k)}$. To avoid redundant forward passes during training, we introduce a memory bank to save and retrievel historical embeddings of $F_{\theta^{(k-1)}}$. The optimal embedding in the memory bank is automatically determined by the early stop operation, where the indicator is average predictive confidence of the sequence. We will introduce the details of the refining technique, knowledge-tuning module and memory bank in the following sections.

\subsection{Refining Technique}
\label{sec:recycling}
\paragraph{Recycling Process} Given the initial residue embedding $\boldsymbol{h}^{(0)} = F_{\theta^{(0)}}(\boldsymbol{x})$, our Knowledge-Design applies a sequence of knowledge-tuning modules to update the residue embedding:
\begin{align}
    \boldsymbol{h}^{(0)} \xrightarrow{f_{\phi^{(1)}}} \cdots \boldsymbol{h}^{(l)} \xrightarrow{f_{\phi^{(l+1)}}} \boldsymbol{h}^{(l+1)} \xrightarrow{f_{\phi^{(l+1)}}} \boldsymbol{h}^{(l+1)} \cdots \xrightarrow{f_{\phi^{(L)}}} \boldsymbol{h}^{(L)}
\end{align}
where $L$ is the maximum number of refinement modules. The predictive probability is obtained by:
\begin{equation}
    \boldsymbol{p}^{(l)} = \texttt{Predict}^{(l)}(\boldsymbol{h}^{(l)})\\
\end{equation}

\paragraph{Virtual MSA} To capture diverse protein knowledge, we sample a set of protein sequences $\{\boldsymbol{s}^{(l), i} \sim \text{Multinomial}(\boldsymbol{p}^{(l)})| 0 \leq i<m\}$ from the predicted probabilities $\boldsymbol{p}^{(l)}$. This set of sequences is called the virtual multiple sequence alignment (MSA). These sequences are fed into pretrained models to obtain the residue embeddings:
 \begin{align}
     \boldsymbol{h}^{(l),i}_{seq} = \mathcal{F}_{seq}(\boldsymbol{s}^{(l), i})\\
     \boldsymbol{h}^{(l),i}_{3d} = \mathcal{F}_{3d}(\boldsymbol{s}^{(l), i}, \boldsymbol{x})\\
 \end{align}
 where $\boldsymbol{x}$ is the 3d coordinates of residues, and $\mathcal{F}_{seq}$ and $\mathcal{F}_{3d}$ are sequence and structure pretrained models, respectively. The sequential embedding $\boldsymbol{h}^{(l),i}_{seq} \in \mathbb{R}^{n, d_{seq}}$ captures the knowledge of the primary sequence, while the structural embedding $\boldsymbol{h}^{(l),i}_{3d} \in \mathbb{R}^{n, d_{3d}}$ captures the knowledge of the 3D protein structure. Together, these features are combined as a unified embedding $\boldsymbol{z}^{(l)}$ through the fusion module:
 \begin{align}
    \boldsymbol{z}^{(l)} = \texttt{Fuse}(\{\boldsymbol{h}^{(l),i}_{seq}\}_{i=1}^m, \{\boldsymbol{h}^{(l),i}_{3d}\}_{i=1}^m, \{\boldsymbol{s}^{(l),i}\}_{i=1}^m, \boldsymbol{p}^{(l)})
\end{align}
which can be further converted as $\boldsymbol{h}^{(l+1)}$ and $\boldsymbol{p}^{(l+1)}$:
\begin{align}
    \boldsymbol{h}^{(l+1)} = f_{\phi^{(l+1)}}(\boldsymbol{z}^{(l)}); \boldsymbol{p}^{(l+1)} = \texttt{Predict}^{(l+1)}(\boldsymbol{h}^{(l+1)})
\end{align}

\paragraph{Confidence-aware updating} We define the confidence vector of a sequence $\boldsymbol{s}$ as the corresponding predictive probability, written as $\boldsymbol{c}_{\boldsymbol{s}}$:
\begin{align}
    \boldsymbol{c}_{\boldsymbol{s}} = [ p_{1, s_1}, p_{2, s_2}, \cdots, p_{n, s_n}  ]^T
\end{align}
Note that $p_{i, s_i}$ represents the predicted probability of the $i$-th amino acid and the predicted residue type is $s_i$. Because some residues are harder to design than others, they may benefit more from refinement. Considering this, we introduce a confidence-aware gated attention mechanism that updates the pre- and post-refinement embeddings based on the predictive confidence of each residue. This allows us to focus more on difficult residues during refinement and improve overall design performance:
\begin{equation}
    \boldsymbol{h}^{(l+1)} \leftarrow \boldsymbol{h}^{(l+1)} \odot \sigmoid( \text{MLP}_1 (\boldsymbol{c}^{(l+1)}_{\boldsymbol{s}'}-\boldsymbol{c}^{(l)}_{\boldsymbol{s}})) + \boldsymbol{h}^{(l)} \odot \sigmoid( \text{MLP}_2 (\boldsymbol{c}^{(l)}_{\boldsymbol{s}}-\boldsymbol{c}^{(l+1)}_{\boldsymbol{s}'}))
\end{equation}
where $\boldsymbol{s} \sim \text{Multinomial}(\boldsymbol{p}^{(l)})$ and $\boldsymbol{s}' \sim \text{Multinomial}(\boldsymbol{p}^{(l+1)})$ are sampled from multimodal distributions. $\sigma$ is the sigmoid function, $\odot$ is element-wise multiplication.

\subsection{Knowledge-tuning Module}
The knowledge-tuning module updates the residue embeddings of well-tuned models to generate more rational protein sequences. As shown in Figure \ref{fig:framework}, the knowledge-tuning module includes a knowledge extractor, a confidence predictor, a fusion layer, and a tuning layer.

\paragraph{Knowledge extractor \& Confidence predictor} As introduced in Sec.\ref{sec:recycling}, the knowledge extractors ($\mathcal{F}_{seq}^{(l)}$ and $\mathcal{F}_{3d}^{(l)}$) are pretrained to extract sequential and structural embeddings from virtual MSAs. The confidence predictor $\texttt{Predict}^{(l)}$ takes the residue embedding $\boldsymbol{h}^{(l)}$ as input and outputs the predictive probability $\boldsymbol{p}^{(l)} \in \mathbb{R}^{n,21}$, which can further be transformed into confidence score $\boldsymbol{c}_{\boldsymbol{s}}^{(l)} \in \mathbb{R}^{n, 1}$.

\paragraph{Fusion layer} The fusion layer combines the sequential and structural embeddings with confidence score to obtain a unified embedding.  Specifically, the structural and sequential MSA embeddings are fused using a confidence-aware gated layer:
\begin{align}
    \boldsymbol{z}^{(l)}  = \sum_{i=1}^m \left[  \text{Embed}(\boldsymbol{s}^{(l),i}) +\text{MLP}_3(\boldsymbol{h}_{seq}^{(l),i})+\text{MLP}_4(\boldsymbol{h}_{3d}^{(l),i}) \right] \odot \sigma(\text{MLP}_5(\boldsymbol{c}_{\boldsymbol{s}}^{(l),i}))
    \label{eq:fusion}
\end{align}

\paragraph{Refinement module} The refinement module is a learnable graph neural network (GNN) that takes $\boldsymbol{z}^{(l)}$ as input node features and $\boldsymbol{e}^{(l)}$ as input edge features. The initial edge features are extracted from the pretrained PiFold model. We use PiGNNs as the refinement module, which consider multi-scale residue interactions and include node updating, local updating, and global updating. The node updating step is as follows:

\begin{equation}
    \label{eq:attention_weight}
    \begin{cases}
        w_{jk} = \text{AttMLP}(\boldsymbol{z}_j^{(l)} || \boldsymbol{e}_{jk}^{(l)} || \boldsymbol{z}_k^{(l)})\\
        a_{jk} = \frac{\exp{w_{jk}}}{\sum_{t \in \mathcal{N}_k}{\exp{w_{tk}}}}\\
        \boldsymbol{v}_j = \text{NodeMLP}(\boldsymbol{e}_{jk}^{(l)} || \boldsymbol{z}_j^{(l)})\\
        \boldsymbol{\hat{z}}_k^{(l)} = \sum_{j \in \mathcal{N}_k}{a_{jk} \boldsymbol{v}_j} 
    \end{cases}
\end{equation}
where $\mathcal{N}_k$ is the neighborhood system of node $k$ and $||$ represents the concatenation operation. At the $l$-th refinement module, $\boldsymbol{z}_j^{(l)}$ is the embeddings of the $j$-th node, $\boldsymbol{e}_{jk}^{(l)}$ is the edge feature between node $j$ and $k$. The edge updating is:
\begin{equation}
   \label{eq:edge}
      \boldsymbol{e}_{jk}^{(l)} \leftarrow \text{EdgeMLP}(\boldsymbol{\hat{z}}_j^{(l)} || \boldsymbol{e}_{jk}^{(l)} || \boldsymbol{\hat{z}}_k^{(l)}) \\
\end{equation}
The global updating applies a gating mechanism to allow the node embeddings interact with the global context. This enables the model to capture long-range dependencies and improve the overall quality of the designed protein sequences:
\begin{equation}
   \label{eq:global}
   \begin{cases}
       \boldsymbol{m}_k = \text{Mean}(\{ \boldsymbol{\hat{z}}_t^{(l)} \}_{t \in \mathcal{B}_k})\\
       \boldsymbol{h}_k^{(l)} \leftarrow \boldsymbol{\hat{z}}_k^{(l)} \odot \sigmoid( \text{GateMLP} (\boldsymbol{m}_k)) \\
   \end{cases}
\end{equation}
where $\mathcal {B}_k$ is the index set of residues belonging to the same protein as residue $k$, $\odot$ is element-wise product operation, and $\sigmoid(\cdot)$ is the sigmoid function.

\subsection{Memory Retrieval}
\vspace{-3mm}

\begin{minipage}{\linewidth}
    \begin{minipage}{0.5\linewidth}
        From Eq.\ref{eq:objective_simplified} we know that $\theta^{(0)}, \phi^{(1)}, \cdots, \phi^{(l)}$ are frozen parameters when optimizing $\phi^{(l+1)}$. Therefore, we can use a memory bank $\mathcal{M}^{(l)}$ to store and retrieve the intermediate embeddings of the $l$-th design model $F_{\theta^{(l)}}(\boldsymbol{x})$ for speeding up the process of optimizing $\phi^{(l+1)}$. As shown in Alg.\ref{alg:memorynet}, the protein embedding $\boldsymbol{h}^{(l)}$ can be retrieved from the memory bank $\mathcal{M}^{(l)}$ without the need for a forward pass, provided that the following conditions are satisfied: (1) the embedding $\boldsymbol{h}^{(l)}$ is already stored in $\mathcal{M}^{(l)}$ and (2) the saved embeddings are consistently obtained from an optimal model $F_{\theta^{(l)}}$. While the first condition is straightforward, the second condition requires the algorithm to automatically determine the optimal $\phi^{(l)}$ and freeze $f_{\phi^{(l)}}$ to ensure that the memorized embeddings are consistent. To determine the optimal $\phi^{(l)}$, we use the average confidence score over the validation set as an indicator and apply the early stopping operation to determine the optimal $\phi^{(l)}$, with a patience of 3.

    \end{minipage}
    \hspace{0.00\linewidth}
    \begin{minipage}{0.5\linewidth}
        \begin{algorithm}[H]
            \caption{Memory Net Framework 
            \\\textbf{Usage}: Retrieve embedding from memory bank without the forward pass.}
            \label{alg:memorynet}
            \begin{small}
            \begin{algorithmic}[1]
                \Require A batch of inputs $\mathcal{H}_{1:b}^{(l)} = \boldsymbol{h}_1^{(l)}|| \boldsymbol{h}_2^{(l)}|| \cdots || \boldsymbol{h}_b^{(l)}$; 
                \Ensure A batch of outputs $\mathcal{H}_{1:b}^{(l+1)} = \boldsymbol{h}_1^{(l+1)} || \boldsymbol{h}_2^{(l+1)} || \cdots || \boldsymbol{h}_b^{(l+1)}$.
                \NoDo
                \NoThen
        
                \State {\color{blue} Step1:} Debatch input data
        
                \State $\{\boldsymbol{h}_1^{(l)}, \boldsymbol{h}_2^{(l)}, \cdots, \boldsymbol{h}_b^{(l)}\} = \text{DeBatch}(\mathcal{H}_{1:b}^{(l)})$;
        
                \State
                
                \State {\color{blue} Step2:} Retrieve embeddings

                \For{$i \in [0, b)$} 
                    \If {$\boldsymbol{h}_i^{(l+1)} \in \mathcal{M}^{(l+1)}$ and $f_{\phi^{(l)}}$ is early stopped}
                        \State $\boldsymbol{h}_i^{(l+1)} = \mathcal{M}^{(l+1)}[i]$
                    \Else
                        \State $\boldsymbol{h}_i^{(l+1)} = \texttt{Refine}^{(l+1)}(\boldsymbol{h}_i^{(l)})$
                    \EndIf
                \EndFor
                \State
        
                \State {\color{blue} Step3:} Batch output
                \State Save $\{\boldsymbol{h}_{1}^{(l+1)}, \boldsymbol{h}_{2}^{(l+1)}, \cdots, \boldsymbol{h}_{n}^{(l+1)}\}$ to $\mathcal{M}^{(l+1)}$
                \State Return $\boldsymbol{h}_1^{(l+1)} || \boldsymbol{h}_2^{(l+1)} || \cdots || \boldsymbol{h}_b^{(l+1)}$
            \end{algorithmic}
        \end{small}
        \end{algorithm}
    \end{minipage}
\end{minipage}

\vspace{-3mm}
\section{Experiments}
\vspace{-3mm}
We evaluate the performance of Knowledge-Design on multiple datasets, including CATH4.2, CATH4.3, TS50, and TS500. We also conduct systematic studies to answer the following questions:
\begin{itemize}[leftmargin=5.5mm]
   \vspace{-2mm}
   \item \textbf{Performance (Q1):} Can Knowledge-Design achieve state-of-the-art accuracy on real-world datasets?
   \vspace{-1mm}
   \item \textbf{Refining technology (Q2):} How much can models gain from different refinement techniques?
   \vspace{-1mm}
   \item \textbf{Knowledge tuning (Q3):}  Which pretrained knowledge is helpful in improving protein design, and how much of a speed boost can the memory bank bring?
   \vspace{-1mm}
   \item \textbf{More analysis (Q4):} How does the Knowledge-Design make a difference on the basis of PiFold?
   \vspace{-1mm}
\end{itemize}
\vspace{-3mm}

\vspace{-3mm}
\subsection{Performance on CATH (Q1)}
\label{sec:exp_cath}
\paragraph{Objective \& Setting} We demonstrate the effectiveness of Knowledge-Design on the widely used CATH \citep{orengo1997cath} dataset. To provide a comprehensive comparison, we conduct experiments on both CATH4.2 and CATH4.3. The CATH4.2 dataset consists of 18,024 proteins for training, 608 proteins for validation, and 1,120 proteins for testing, following the same data splitting as GraphTrans \citep{ingraham2019generative}, GVP \citep{jing2020learning}, and PiFold \citep{gao2023pifold}. The CATH4.3 dataset includes 16,153 structures for the training set, 1,457 for the validation set, and 1,797 for the test set, following the same data splitting as ESMIF \citep{hsu2022learning}. The model is trained up to 20 epochs using the Adam optimizer on an NVIDIA V100. The batch size and learning rate used for training are 32 and 0.001, respectively. To evaluate the generative quality, we report perplexity and median recovery scores on short-chain, single-chain, and all-chain settings.

\paragraph{Baselines} To evaluate the performance of Knowledge-Design, we compare it with recent graph models, including StructGNN, StructTrans \citep{ingraham2019generative}, GCA \citep{tan2022generative}, GVP \citep{jing2020learning}, GVP-large, AlphaDesign \citep{gao2022alphadesign}, ESM-IF \citep{hsu2022learning}, ProteinMPNN \citep{dauparas2022robust}, and PiFold \citep{gao2023pifold}, as most of them are open-source. To ensure a fair and reliable comparison, we reproduce StructGNN, StructTrans, GCA, GVP, AlphaDesign, ProteinMPNN, and PiFold under the same data splitting as ours on the CATH 4.2 dataset. To provide a head-to-head comparison with ESMIF, we retrain our model on the CATH4.3 dataset following the same data splitting as ESMIF.

\begin{table}[h]
   \centering
   \caption{Results comparison on the CATH dataset. All baselines are reproduced under the same code framework, except ones marked with $\dagger$. We copy results of GVP-large and ESM-IF from their manuscripts \citep{hsu2022learning}. The \textbf{best} and \underline{suboptimal} results are labeled with bold and underline.}
   \label{tab:results_cath}
   \resizebox{1.0 \columnwidth}{!}{
   \begin{tabular}{lcccccccc}
   \hline
   \multirow{2}{*}{Model} & \multicolumn{3}{c}{Perplexity $\downarrow$} & \multicolumn{3}{c}{Recovery \% $\uparrow$} & \multicolumn{2}{c}{CATH version}\\
   & Short   & Single-chain  & All  & Short   & Single-chain   & All  & 4.2  & 4.3\\ \hline
   StructGNN              &  8.29   &   8.74        &  6.40    &   29.44      &   28.26             &  35.91   & $\checkmark$ \\
   GraphTrans             &    8.39            &    8.83         &    6.63          &   28.14        &     28.46          &    35.82   & $\checkmark$ \\
   GCA                    &  7.09     &  7.49  &  6.05  & 32.62 & 31.10 & 37.64 & $\checkmark$\\
   GVP                    &  7.23       &   7.84           &  5.36    &   30.60      &     28.95           &  39.47   & $\checkmark$ \\
   GVP-large$^\dagger$              &  7.68       &    \underline{6.12}           &  6.17    &   32.6      &     \underline{39.4}           &   39.2  & & $\checkmark$ \\
   AlphaDesign                &  7.32       &    7.63           &  6.30    &  34.16       &    32.66            &  41.31  & $\checkmark$  \\
   ESM-IF$^\dagger$         &   8.18      &    6.33           &  6.44    &   31.3      &     38.5           &  38.3   & & $\checkmark$ \\
   ProteinMPNN            &  {6.21}       &   6.68            &  {4.61}    &  {36.35}       &    34.43            &  45.96   & $\checkmark$ \\
   PiFold              &  \underline{6.04}       &   {6.31}            &  \underline{4.55}    &   \underline{39.84}      &     {38.53}           &  \underline{51.66}   & $\checkmark$ \\ \hline
   Knowledge-Design (Ours)  &  \textbf{5.48}       &    \textbf{5.16}           &  \textbf{3.46}    &   \textbf{44.66}      &    \textbf{45.45}            &  \textbf{60.77}   & $\checkmark$ \\ 
   \hline
   \end{tabular}}
   \vspace{-3mm}
\end{table}

\paragraph{Results \& Analysis} Based on the results presented in Table \ref{tab:results_cath}, we can see that Knowledge-Design consistently achieves state-of-the-art performance on different settings, with significant improvements over previous models. Specifically, we observe the following:
(1) Knowledge-Design is the first model to exceed 60\% recovery on both CATH4.2 and CATH4.3, demonstrating its superior ability in generating protein structures. (2) On the full CATH4.2 dataset, Knowledge-Design achieves a perplexity of 3.46 and a recovery of 60.77\%, outperforming the previous state-of-the-art model PiFold by 23.95\% and 9.11\%, respectively. Furthermore, Knowledge-Design achieves a recovery improvement of 4.82\% and 6.92\% on the short and single-chain settings, respectively. (3) Knowledge-Design also achieves similar improvements when extending to the CATH4.3 dataset, further validating its effectiveness and generalizability. Overall, these results demonstrate the superior performance and potential of Knowledge-Design in protein design, and suggest that it could be a valuable tool for advancing protein engineering and drug design.

\subsection{Performance on TS50 and TS500 (Q1)}
\paragraph{Objective \& Setting} To provide a more comprehensive evaluation and demonstrate the generalizability of Knowledge-Design, we also evaluate it on two standard protein benchmarks, TS50 and TS500. These datasets contain 50 and 500 proteins, respectively, and are widely used for evaluation. In addition to graph-based models, we also include MLP- and CNN-based methods as baselines to provide a more comprehensive comparison.

\begin{table}[h]
   \small
   \caption{Results on TS50 and TS500. All baselines are reproduced under the same code framework, except ones marked with $\dagger$, whose results are copied from their manuscripts. The \textbf{best} and \underline{suboptimal} results are labeled with bold and underline.}
   \label{tab:TS_results}
   \centering
   \resizebox{1.0 \columnwidth}{!}{
   \begin{tabular}{llcccccc}
   \hline
   \multirow{2}{*}{Group}       & \multirow{2}{*}{Model}  & \multicolumn{3}{c}{TS50} & \multicolumn{3}{c}{TS500} \\ \cmidrule(lr){3-5} \cmidrule(lr){6-8} 
                                &                        & Perplexity $\downarrow$    & Recovery $\uparrow$  &  Worst $\uparrow$ & Perplexity $\downarrow$  & Recovery $\uparrow$  & Worst $\uparrow$ \\ \hline
   \multirow{3}{*}{\rotatebox{90}{MLP}}        
   & SPIN $^\dagger$                  &                &     30.30        &              &          &  30.30 &             \\
   & SPIN2  $^\dagger$                &                &     33.60        &     &         &    36.60            \\
   & Wang's model $^\dagger$          &                &     33.00        &              &     & 36.14         \\ \hline
   \multirow{3}{*}{\rotatebox{90}{CNN}}            
   & SPROF  $^\dagger$                &                &     39.16        &              &           &      40.25         &           \\
     & ProDCoNN  $^\dagger$             &                &    40.69         &              &           &     42.20          &           \\
      & DenseCPD  $^\dagger$             &                &     50.71        &              &           &    55.53           &           \\ \hline
   \multirow{7}{*}{\rotatebox{90}{Graph}} & StructGNN              &    5.40            &   43.89          &    26.92          &   4.98        &    45.69           &   \underline{0.05}        \\
    & GraphTrans       &   5.60  & 42.20 & 29.22 &  5.16 & 44.66 & 0.03  \\
    & GVP   & 4.71  & 44.14 & 33.73 & 4.20 & 49.14 & \textbf{0.09}\\
    & GCA       &     5.09           &    47.02         &     28.87         &   4.72        &    47.74           &   0.03        \\
    & AlphaDesign       & 5.25 & 48.36 & 32.31 & 4.93 & 49.23 & 0.03 \\
    & ProteinMPNN            & {3.93}  & {54.43} & {37.24} & {3.53} & {58.08} & 0.03\\
    & PiFold               &   \underline{3.86}             &    \underline{58.72}         &    \underline{37.93}          &   \underline{3.44}        &    \underline{60.42}           &    0.03       \\ 
    & Knowledge-Design (Ours) &   \textbf{3.10}             &    \textbf{62.79}         &    \textbf{39.31}          &   \textbf{2.86}        &    \textbf{69.19}           &    0.02       \\ \hline
   \end{tabular}}
\end{table}

\paragraph{Results and Analysis} Experimental results are shown in Table.\ref{tab:TS_results}, where Knowledge-Design significantly outperforms previous baselines on all benchmarks. We observe that: (1) On the TS50 dataset, Knowledge-Design achieves a perplexity of 3.10 and a recovery rate of 62.79\%, outperforming the previous state-of-the-art model, PiFold, by 19.69\% and 4.07\%. (2) On the TS500 dataset, Knowledge-Design achieves a perplexity of 2.86 and a recovery rate of 69.19\%, outperforming PiFold by 16.86\% and 8.77\%. (3) Notably, Knowledge-Design is the first model to exceed 60\% and 65\% recovery on the TS50 and TS500 benchmarks, respectively.

\subsection{Refining technology (Q2)}
\paragraph{Objective \& Setting}  We conduct ablation studies to investigate the effects of virtual MSA, recycling, and the confidence-aware tuning module. We follow the same experimental setting as in Section \ref{sec:exp_cath} and report the results on the CATH dataset. Specifically, we vary the number of virtual MSA and recycling times from 1 to 3, and remove the confidence-aware tuning module by replacing the confidence score with a constant value of 1.0. We also compare the training time with and without using the memory bank.

\begin{table}[h]
   \resizebox{1.0 \columnwidth}{!}{\begin{tabular}{ccccccccccc}
   \hline
    \multicolumn{3}{c}{Config}                            & \multicolumn{3}{c}{Perplexity $\downarrow$}                        & \multicolumn{3}{c}{Recovery \% $\uparrow$}       & \multicolumn{2}{c}{Training time (per epoch)$\downarrow$}                  \\
    w/o confidence                   & \multicolumn{1}{c}{MSA} & \multicolumn{1}{c}{Recycle} & Short                & Single-chain         & All                  & Short                & Single-chain         & All      & w/o memory & w memory     \\ \hline
                   & 1                       & 1                           &    5.54                  &    5.39                  &  3.59                    &    42.58                  &     42.74                 &  58.39           & 20min & 70 min        \\
                   & 1                       & 2                           &     5.52                 &    5.31                  &  3.52                    &    44.72                  &      44.19                &   59.72            & 40min     & 140min  \\
                   & 1                       & 3                           & 5.46 & 5.17 & 3.48 & 43.91 & 44.16 & 60.34 & 60min  & 210min\\ \hline
                   & 1                       & 1                           &    5.54                  &    5.39                  &  3.59                    &    42.58                  &     42.74                 &  58.39           & 20min & 70 min        \\
                   & 2                       & 1                           &    5.55                  &    5.42                  &     3.56                 &     42.72                 &   42.16                   & 58.62  & 33min         & 83min          \\
                   & 3                       & 1                           &  5.57                    &    5.42                  &     3.56                 &   42.94                   &   43.23                   &   58.71         & 45min     & 95min     \\ \hline
                   & 2                       & 2                           &  5.49                    &     5.22                 &    3.49                  &     44.76                 &     45.90                 &  59.84             &   65min   & 165min  \\
                   & 2                       & 3                           & 5.48 & 5.16 & 3.46 & 44.66 & 45.45 & 60.77 & 100min & 250min\\ 
   \checkmark                & 2                       & 3                           & 5.50 & 5.24 & 3.52 & 43.88 & 44.08 & 59.64 & -- & -- \\ \hline
   \end{tabular}}
   \caption{Ablation of refining technology. "w/o confidence means" replacing the confidence score as a consisent value 1.0.  "w memory" and "w/o memory" indicates whether using the memory-retrievel mechanism or not. The training time is measured on an NVIDIA V100. }
   \label{tab:refine}
\end{table}

\vspace{-3mm}
\paragraph{Results and Analysis} Ablation studies about MSA, recycling and confidence embedding are presented in Table \ref{tab:refine}. We conclude that:(1) Recycling has a more significant impact on performance than virtual MSA. When increasing the recycling from 1 to 3 while keeping the number of virtual MSAs constant at 1, the recovery rate on the full dataset improves by 1.95\%, from 58.39\% to 60.34\%. In contrast, increasing the number of virtual MSAs from 1 to 3 only results in a 0.32\% improvement. (2) The confidence-aware tuning module make an non-trival improvement by 0.78\%, 1.37\%, and 1.13\% on the short, single-chain, and full datasets, respectively. (3) Increasing the number of virtual MSAs and recycling times leads to higher computational overhead during training. Therefore, we introduce a memory retrieval mechanism that saves more than 50\% of the training time in all cases. (4) Based on above analysis, we conclude that the importance order of the three components is as follows: recycling $>$ confidence-aware tuning module $>$ virtual MSA.

\subsection{Pretrain Knowledge(Q3)}
\paragraph{Objective \& Setting} By deleting the corresponding pre-trained embedding in Eq.\ref{eq:fusion}, we investigate how much performance gain the model can achieve from different pre-trained models, including ESM-650M \citep{meier2021language,lin2022language}, ESM-IF \citep{hsu2022learning}, and GearNet \citep{zhang2022protein}. The experimental settings keep the same as in Section \ref{sec:exp_cath}. 

\begin{table}[h]
   \centering
   \caption{Ablation study on multimodal knowledge. We investigate how much performance gain the model can obtain from different pre-trained models, where ESM+ESMIF provides the best performance. }
   \label{tab:results_knowledge}
   \resizebox{0.8 \columnwidth}{!}{\begin{tabular}{lcccccc}
   \toprule
   \multirow{2}{*}{Model} & \multicolumn{3}{c}{Perplexity $\downarrow$} & \multicolumn{3}{c}{Recovery \% $\uparrow$} \\
   & Short   & Single-chain  & All  & Short   & Single-chain   & All  \\ \midrule
   PiFold              &  {6.04}       &   {6.31}            &  {4.55}    &   {39.84}      &     {38.53}           &  {51.66}   \\
   Knowledge-Design (GearNet) &  6.66       &   6.89            &  4.96    &   38.72      &     38.02           &  50.43   \\ 
   Knowledge-Design (ESM) &  \underline{6.05}       &   \underline{5.29}            &  \underline{3.90}   &   \underline{43.32}      &    \underline{46.30}            &  \underline{57.38}    \\ 
   Knowledge-Design (ESMIF) &  6.15       &   6.51            &  4.18    &   38.79      &     39.71           &  54.52    \\ 
   Knowledge-Design (ESMIF+ESM) & \textbf{5.48} & \textbf{5.16} & \textbf{3.46} & \textbf{44.66} & \textbf{45.45} & \textbf{60.77}    \\ \bottomrule
   \end{tabular}}
   \vspace{-3mm}
\end{table}

\paragraph{Results \& Analysis} We present results on Table.\ref{tab:results_knowledge}. We observe that: (1) The knowledge of ESMIF and ESM pre-trained models contributes to improving the performance, with ESM providing a larger improvement than ESMIF. Specifically, the knowledge of ESMIF and ESM results in a 2.86\% and 5.72\% improvement, respectively. In contrast, the knowledge of GearNet does not contribute to the improvement. (2) The best recovery rate is achieved when the model combines the knowledge of ESM and ESMIF, resulting in a 9.11\% improvement. Notably, the improvement is not linear, as the combination of ESMIF and ESM provides a larger improvement than the sum of their individual contributions (i.e., 9.11\% > 2.86\% + 5.72\%). These results highlight the importance of selecting the appropriate pre-trained models for protein structure refinement and demonstrate the effectiveness of combining multiple sources of knowledge to achieve better performance.

\subsection{More Analysis(Q4)}
\begin{figure}[h]
   \centering
   \includegraphics[width=5in]{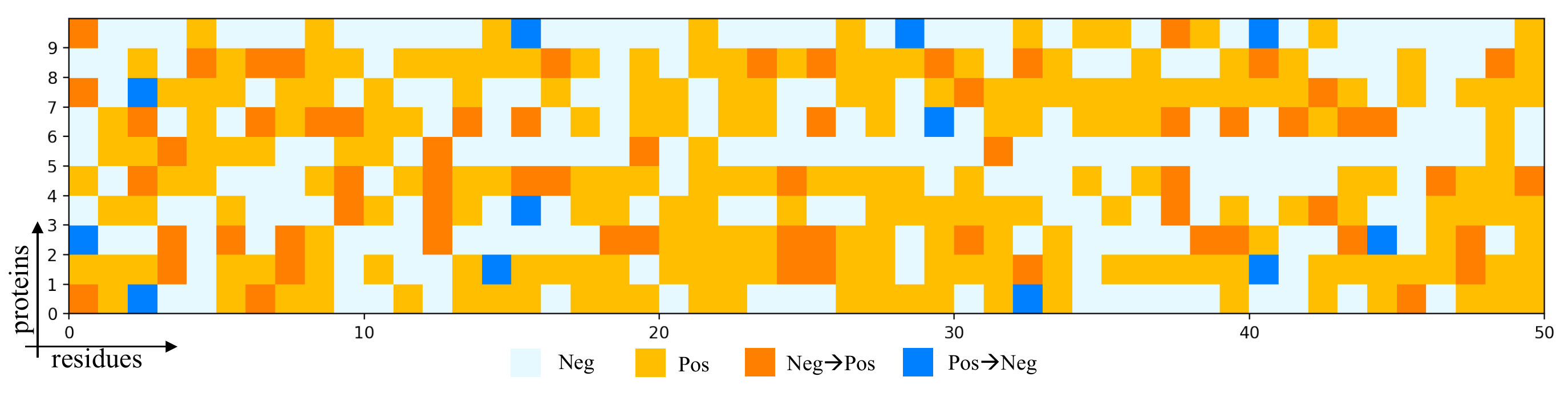}
   \vspace{-3mm}
   \caption{Recovery states. The light blue cell indicates a negative residue designed by PiFold and differs from the reference one, while the light orange cell indicates a positive residue that matches the reference one. The darker blue cells indicate where Knowledge-Design wrongly converted positive residues into negative ones, while the darker orange cells indicate where Knowledge-Design corrected negative residues to positive ones. }
   \label{fig:refine_details}
   \vspace{-3mm}
\end{figure}

\vspace{-3mm}
\paragraph{Recovery States} We randomly selected 10 proteins from the CATH4.2 dataset test set and designed their sequences using PiFold. The sequences were then refined using Knowledge-Design, and the recovery states of the designed residues were visualized in Fig. \ref{fig:refine_details}. Our results show that Knowledge-Design tends to make more positive corrections than negative corrections, with positive corrections occurring mostly in adjacent locations to initially positive residues. This suggests that the model learns the local consistency of the protein structure and can automatically correct incorrectly designed residues that violate this consistency.

\vspace{-3mm}

\begin{minipage}{\linewidth}
   \begin{minipage}{0.57\linewidth}
      \paragraph{Distribution comparison} Fig. \ref{fig:compare_distribution} shows the confidence distributions of positive and negative residues generated by PiFold and Knowledge-Design on the CATH4.2 test set. Positive residues tend towards a confidence of 1.0, while negative residues have mostly below 0.6 confidence, indicated by different colors. Our results demonstrate that Knowledge-Design produces positive residues with higher confidence compared to PiFold, while also reducing the number of negative residues. This suggests that Knowledge-Design can convert low-confidence positive residues to high-confidence ones and correct negative residues as positive ones. 
   \end{minipage}
   \hspace{0.00\linewidth}
   \begin{minipage}{0.43\linewidth}
      \begin{figure}[H]
         \centering
         \includegraphics[width=2.3in]{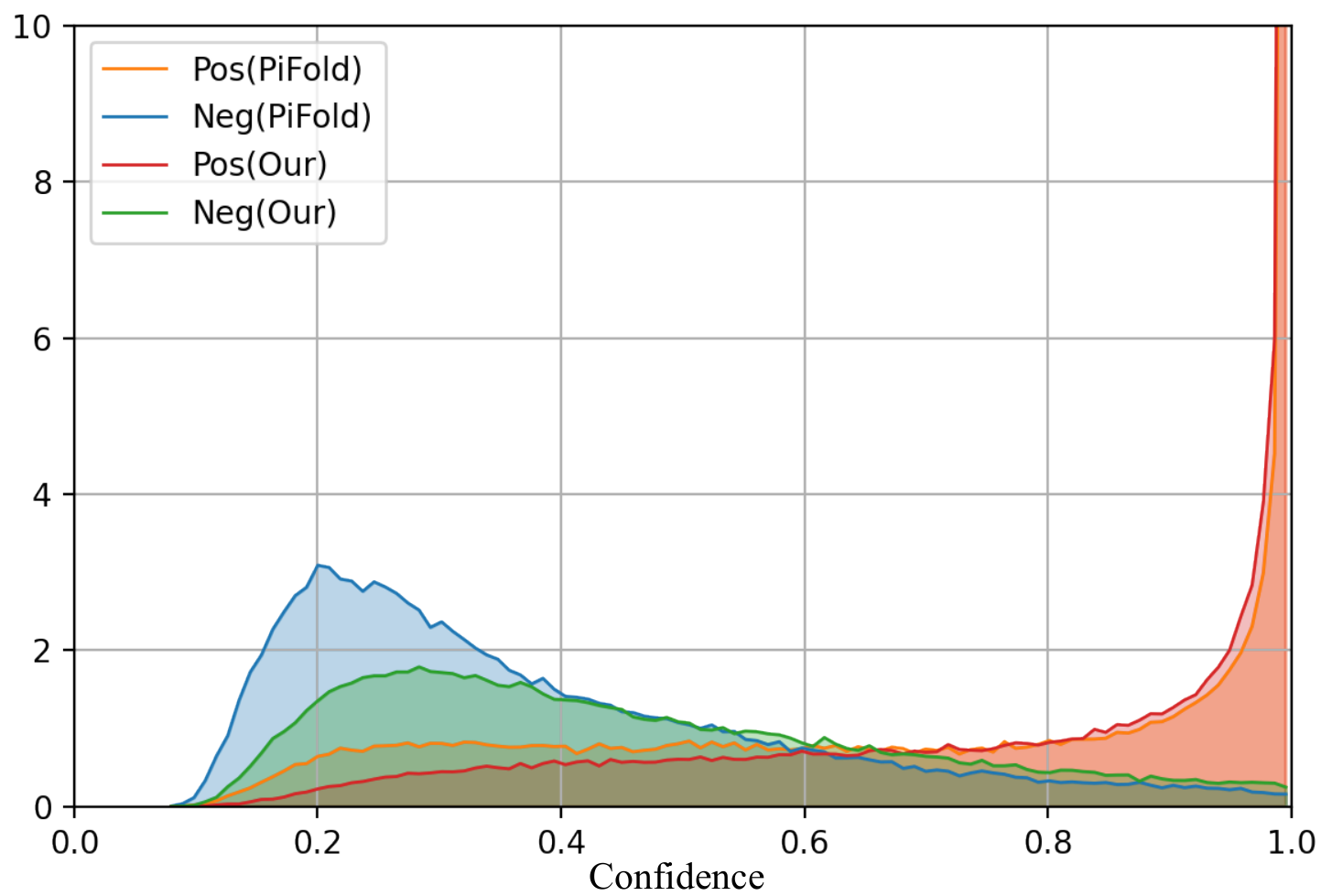}
         \caption{Confidence distributions. }
         \label{fig:compare_distribution}
      \end{figure}
   \end{minipage}
\end{minipage}

\begin{figure}[h]
   \centering
   \includegraphics[width=4in]{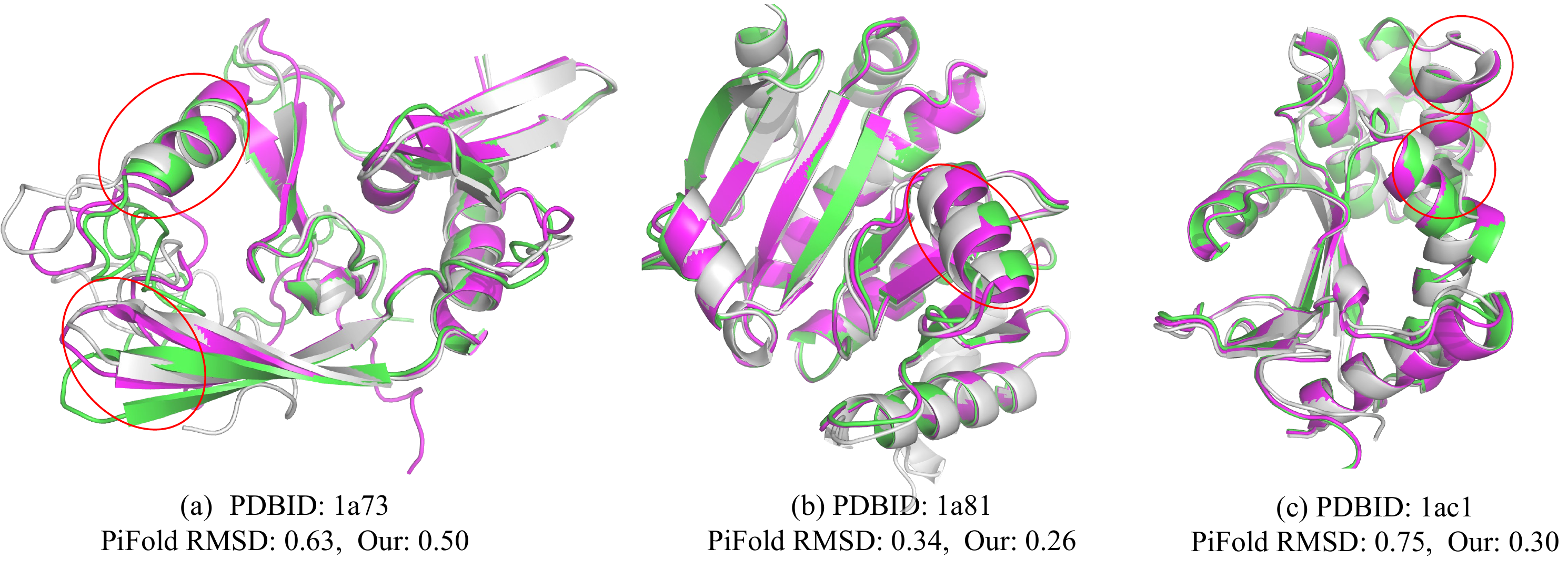}
   \vspace{-3mm}
   \caption{Comparing the designed proteins. The green structures are reference ones, while the gray and purple structures are designed by PiFold and Knowledge-Design, respectively. We use red circles to highlight the regions where Knowledge-Design produces more similar structures to the reference ones than PiFold.}
   \label{fig:design_example}
   \vspace{-3mm}
\end{figure}

\vspace{-3mm}
\paragraph{Compare structures} In Fig.\ref{fig:design_example}, we use ESMFold\citep{lin2022language} to generate protein structures from designed sequences and comapring the designed proteins of PiFold and Knowledge-Design against the reference ones.  We observe that the designed structures of Knowledge-Design are more similar to the reference ones than that of PiFold. Specifically, Knowledge-Design achieves 15.9\%, 35.3\%, and 60\% improvement in root mean square deviation (RMSD) on the 1a73, 1a81, and 1ac1 proteins, respectively. These results demonstrate that Knowledge-Design can generate proteins that are structurally more similar to the reference ones compared to PiFold.

\vspace{-3mm}

\vspace{-3mm}
\section{Conclusion\&Limitation} 
\vspace{-3mm}
We propose Knowledge-Design, a novel method that iteratively refines low-confidence residues using common protein knowledge extracted from pretrained models. Knowledge-Design is the first model that achieves 60+\% recovery on CATH4.2, CATH4.3, TS50, and TS500, demonstrating its effectiveness and generalizability. However, the proposed method has not yet been verified through wet experiments in real applications, and this will be a direction for future work.

\bibliographystyle{plainnat} 

\bibliography{main}

\clearpage
\appendix
\section{Appendix}
\paragraph{Compare to LMDesign} In parallel with our work, we are observing another exciting project called LMDesign \citep{zheng2023structure}, which was recently published at ICML as an oral presentation. LMDesign aims to use the pre-trained ESM model to improve protein design. However, there are several differences between our knowledge-Design and LMDesign. 
\begin{itemize}
    \item \textbf{More comprehensive}: We enhance protein design by fusing multimodal knowledge from pre-trained models, including both structural and sequential information, while LMDesign only uses single-modal information. Our experiments demonstrate that combining these modalities leads to nontrivial improvements, as shown in Table \ref{tab:results_knowledge}
    \item \textbf{More efficient}: We introduce the memory-retrieval mechanism to save more than 50\% of the training time, while LMDesign does not use this mechanism.
    \item \textbf{Novel modules}: We introduce confidence-aware recycling techniques as well as virtual MSA to boost the model performance.
    \item \textbf{More effective}:  Overall, our model outperforms LMDesign by 5.12\% on the CATH4.2 dataset.
\end{itemize}

\begin{table}[h]
  \centering
  \caption{Results comparison on the CATH dataset. All baselines are reproduced under the same code framework, except ones marked with $\dagger$. We copy results of GVP-large and ESM-IF from their manuscripts \citep{hsu2022learning}. The \textbf{best} and \underline{suboptimal} results are labeled with bold and underline.}
  \label{tab:cmp_lmdesign}
  \resizebox{1.0 \columnwidth}{!}{
  \begin{tabular}{lcccccccc}
  \hline
  \multirow{2}{*}{Model} & \multicolumn{3}{c}{Perplexity $\downarrow$} & \multicolumn{3}{c}{Recovery \% $\uparrow$} & \multicolumn{2}{c}{CATH version}\\
  & Short   & Single-chain  & All  & Short   & Single-chain   & All  & 4.2  & 4.3\\ \hline
  StructGNN              &  8.29   &   8.74        &  6.40    &   29.44      &   28.26             &  35.91   & $\checkmark$ \\
  GraphTrans             &    8.39            &    8.83         &    6.63          &   28.14        &     28.46          &    35.82   & $\checkmark$ \\
  GCA                    &  7.09     &  7.49  &  6.05  & 32.62 & 31.10 & 37.64 & $\checkmark$\\
  GVP                    &  7.23       &   7.84           &  5.36    &   30.60      &     28.95           &  39.47   & $\checkmark$ \\
  GVP-large$^\dagger$              &  7.68       &    \underline{6.12}           &  6.17    &   32.6      &     {39.4}           &   39.2  & & $\checkmark$ \\
  AlphaDesign                &  7.32       &    7.63           &  6.30    &  34.16       &    32.66            &  41.31  & $\checkmark$  \\
  ESM-IF$^\dagger$         &   8.18      &    6.33           &  6.44    &   31.3      &     38.5           &  38.3   & & $\checkmark$ \\
  ProteinMPNN            &  {6.21}       &   6.68            &  {4.61}    &  {36.35}       &    34.43            &  45.96   & $\checkmark$ \\
  PiFold              &  \underline{6.04}       &   {6.31}            &  {4.55}    &   \underline{39.84}      &     {38.53}           &  {51.66}   & $\checkmark$ \\ 
  LMDesign & 6.77 & 6.46 & \underline{4.52} & 37.88 & \underline{42.47} & \underline{55.65} & $\checkmark$\\
  \hline
  Knowledge-Design (Ours)  &  \textbf{5.48}       &    \textbf{5.16}           &  \textbf{3.46}    &   \textbf{44.66}      &    \textbf{45.45}            &  \textbf{60.77}   & $\checkmark$ \\ 
  \hline
  \end{tabular}}
  \vspace{-3mm}
\end{table}

\end{document}